\documentstyle[aps,pra,epsfig,twocolumn]{revtex}

\def\be{\begin{equation}}
\def\ee{\end{equation}}
\def\bea{\begin{eqnarray}}
\def\eea{\end{eqnarray}}
\def\bma{\begin{mathletters}}
\def\ema{\end{mathletters}}
\def\bi{\begin{itemize}}
\def\ei{\end{itemize}}
\def\tr{{\rm tr}}

\def\C{\hbox{$\mit I$\kern-.7em$\mit C$}}

\newcommand{\one}{\mbox{$1 \hspace{-1.0mm}  {\bf l}$}}
\newcommand{\ket}[1]{ | \, #1  \rangle}
\newcommand{\bra}[1]{ \langle #1 \,  |}
\newcommand{\proj}[1]{\ket{#1}\bra{#1}}

\begin{document}
\draft

\title{Characterization of Distillable and Activable States using Entanglement Witnesses}

\author{B. Kraus$^1$, M. Lewenstein$^2$,  and J.I. Cirac$^1$}

\address{$^1$Institut f\"ur Theoretische Physik, Universit\"at Innsbruck,A-6020
Innsbruck, Austria\\
$^2$Institut f\"ur Theoretische Physik, Universit\"at Hannover, Hannover, Germany}

\date{\today}

\maketitle

\begin{abstract}

We introduce a formalism that connections entanglement witnesses and the distillation and activation properties of a state. We apply this formalism to two cases: First, we rederive the results presented in {\bf quant-ph/0104095} \cite{Eg01} by Eggeling et al., namely that on copy of any bipartite state with non--positive partial transpose (NPPT) is either distillable, or activable. Second, we show that there exist three--partite NPPT states, with the property that two copies can neither be distilled, nor activated. 

\end{abstract}

\narrowtext

\section{Introduction}

Entanglement is one of the most fascinating features in Quantum Mechanics. 
It has been shown that maximally entangled states of two parties can be 
used in many applications of Quantum Information \cite{QI}. For instance one can teleport 
a state of a particle to another particle \cite{Bennett_tel}, which is spatially separated from
it. It was also shown how to use entangled state to send secret messages from one place to another 
\cite{Ek91}. In most of those proposals one needs pure maximally entangled states. In reality, 
however, the states which are produced in the laboratories are, due to the interaction with 
the environment mixed. 

It was shown by  Bennett et al. \cite{Be96}, Deutsch et al. \cite{De96}, 
and by Gisin \cite{Gi96}, how to obtain, out of a some copies of an entangled mixed state, 
pure maximally entangled 
states, using only local operations and classical communication\cite{Li98,Ke98}. 
This process is called distillation. Later on the Horodecki family proved that 
any entangled state of two qubits (two--level) systems
can be distilled to a maximally entangled state \cite{Hodist22}. 
They also showed that a necessary condition for 
distillability is that the partial transpose (\ref{parttra}) of the density operator must 
be non--positive semidefinite \cite{Hodist}.
In higher dimensions, however, there exist states, fulfilling this necessary condition for distillation, 
but it is not possible to transform some copies of this state into a maximally entangled state \cite{Du99,Di00}. 
Then there is still the possibility to distill some entanglement, if one allows Alice and Bob to
share, in addition to there states an entangled state, whose partial transpose is positive (PPTES) \cite{Ho99}.
This process is called activation, since the entanglement contained in the copies of the original state is activated by a PPTES.

Recently, it has been shown \cite{Eg01} that one copy of any bipartite state can always be either 
distilled or activated by a PPTES \cite{Sh01}. We will show that this is not the case if we consider systems composed of more than two subsystems. That is we give an example of a three--partite state, where 
one (even two) copy of it can neither be distilled, nor activated. We will introduce a formalism, 
which allows to connect the problem of entanglement witnesses (EW), which are observables that 
allow to detect entangled states, to the problem of distillation and activation, for arbitrary states \cite{note1}.

This paper is divided into $4$ sections. In Section II we introduce our notation and summarize some known results concerning separability. There we will also generalize the notion of entanglement witnesses (EW) 
and some results concerning EW's \cite{Le00,Le01} to more than two systems \cite{Le00,Ka01,HoEwn}. 
Then we review the results on distillation and activation of entanglement \cite{Hodist22}. 
The technical details, concerning EW's, but also the notion of completely positive maps (CPM) 
are written in an appendix. There we also reformulate the problem of distillation and activation 
of entanglement in terms of CPM's. The reason for that is that the main results of this paper can 
be understood without those technical details, but we need them to prove our statements. 
Section III is divided into two parts. In the first part we consider a density operator, 
$\rho$ which describes the state of a system composed of two subsystems. For an arbitrary number, 
say $N$, of copies of this state we define an operator $W_{\rho^{\otimes N}}$. Then we show that 
those $N$ copies of the state can be distilled iff $W_{\rho^{\otimes N}}$ is no EW; the entanglement 
of those $N$ copies can be activated via a PPTES iff $W_{\rho^{\otimes N}}$ is a special EW, 
namely a non--decomposable EW (NDEW). Furthermore, if the entanglement of a state can be distilled, then the introduced formalism gives us a distillation protocol. If the entanglement can be activated, then we know which PPTES activates it. In the second part of Section III we generalize those
results to density operators describing 
more than two parties, by concentrating on three systems. Section IV contains two applications of the formalism developed. First we rederive the above explained result presented in \cite{Eg01} in a simple 
manner. Second, we present an example of a three--partite state, whose partial transpose is
not positive semidefinite, but, nevertheless one copy of the state can neither be distilled, nor 
activated. We then show that even if we consider two copies of some of those states the entanglement 
can neither be distilled, nor activated. Section V contains a summary of the results.

\section{Notation and Review}
The aim of this section is two--fold. On the one hand we introduce our notation and on the other hand we summarize some known facts which we use to prove the main results of this paper. In the first subsection we recall the notion of separability. Then we generalize the results concerning entanglement witnesses \cite{Le00,Le01} to more than two parties\cite{HoEwn}. In the last section we recall the notion of distillation and activation of entanglement. As mentioned before, the technical details concerning EW's, the notion of CPM's and the connection between CPM's and the distillation and activation problem can be found in the appendix. 
                                              
Throughout this paper we denote by $\{\ket{1},\ldots\ket{d}\}$ the computational basis in $\C^d$. Whenever we consider two or more
systems, $A,B,\ldots$ we use the notation $\ket{i}_A\ket{j}_B=\ket{i,j}_{AB}$ and if it is clear to which Hilbert spaces the states belong to then we omit the subscripts. For instance we write the (unnormalized) maximally entangled state in $\C^{d_1}\otimes \C^{d_2}$ as,
\bea
\label{maxent}
\ket{\Phi_d}=\sum_{k=1}^{d}\ket{k,k},
\eea
where $d=\mbox{min}\{d_1,d_2\}$. In the following the superscript $T$ denotes the 
transposition in the computational basis. By ${\cal B}(\cal H)$ we denote the Hilbert space of 
bounded operators acting on the Hilbert space ${\cal H}$. Furthermore we denote by $\one_n$ the 
identity matrix acting on a $n$--dimensional Hilbert space. Most of the paper will deal with 
composite systems. In this case the Hilbert spaces of the spatially separated systems will 
be denoted by ${\cal H}_X$, with $\mbox{dim}({\cal H}_X)=d_X$. We will only consider 
the nontrivial situations when $d_X\ge 2$. 
Whenever we consider the situation where Alice, Bob, Charly, $\ldots$ have more than
one particle we will denote them by $A_1,A_2,\ldots A_n$, $B_1,B_2,\ldots B_n$, $C_1,C_2,\ldots C_n$, $\ldots$ respectively. The corresponding total Hilbert space of each party is then ${\cal H}_X={\cal H}_{X_1}\otimes \ldots {\cal H}_{X_n}$, for $X\in \{A,B,C,\ldots\}$. We will also use the notation ${\cal H}_i={\cal H}_{A_i}\otimes \ldots \otimes {\cal H}_{X_i}$, for $i\in
\{1,\ldots ,n\}$. Capital letters as sub--or superscript will indicate on which system an operator acts on, e.g. $O^A$ denotes an operator acting an ${\cal H}_A$. For simplicity we will not normalize the states which we consider.  

Let us now recall some facts concerning separability, entanglement witnesses, and distillation and 
activation properties. Throughout this section we consider a density operator, $\rho$, 
describing the state of several, say $N$, spatially separated systems. The Hilbert space on 
which $\rho$ acts is ${\cal H}={\cal H}_A \otimes \ldots \otimes {\cal H}_Z$. 

\subsection{Separability}
A state, $\rho$, is called fully separable if it can be prepared, using only local operations and classical communication (LOCC), out of a product state, e.g. $\ket{0,\ldots,0}$. Equivalently,  a state, $\rho$, is fully separable iff it can be written as
\bea
\label{sep} \rho=\sum_{i} p_i
\ket{a_i}_A\bra{a_i}\otimes \ldots \otimes
\ket{z_i}_Z\bra{z_i},
\eea
where $p_i\geq 0$ and the $\ket{x_i}$ belong to the Hilbert space of particle $X$. If $\rho$ cannot be written as (\ref{sep}), it is inseparable (entangled). In what follows we simply call a state or a map separable if it is fully separable. It was shown by Peres \cite{Pe96} and the Horodecki family \cite{Ho96} that a density matrix, $\rho$, which describes the state of two qubits (${\cal H}=\C^2\otimes \C^2$), or one qubit and a three level system (${\cal H}=\C^2\otimes \C^3$) is separable iff its partial transpose is positive semidefinite. Here, the partial transpose of an hermitian operator, $O$, with respect to system $Y$ in the computational basis \cite{note0} is defined as:
\bea
\label{parttra}
O^{T_Y}=\sum_{i,j=1}^{d_Y}\phantom{.}_y \bra{i} O \ket{j}_y \ket{j}_y \bra{i}.
\eea
From the condition (\ref{sep}) it can be easily seen that a state, $\rho$, is separable iff $\rho^{T_Y}$ is separable, for any system $Y$. Thus, it is clear that the partial transpose of a separable state is positive semidefinite. In the following we call a state, $\rho$, $Y$--PPT if its partial transpose with respect to system $Y$ is a positive semidefinite operator. Otherwise we call it $Y$--NPPT. If a state has a positive (non--positive) semidefinite partial transpose with respect to all systems then we call it simply PPT (NPPT). If we consider, for instance, only two
systems, we have $\rho^{T_B}=(\rho^{T_A})^T$ and therefore $\rho$ is $A$-NPPT iff it is $B$-NPPT; we call such a state NPPT. Throughout the paper
we use the fact that $\tr(\rho X^{T_A})= \tr(\rho^{T_A} X)$.

Note that in higher dimensions, (${\cal H}=\C^n \otimes\ \C^m$), where $n+m>5$, or if the state describes a system composed of more than two
parties, there exist entangled states whose partial transposes are positive semidefinite operators (PPTES)\cite{Ho97}. Thus, in this case the positivity of the partial transpose is no longer a sufficient, but only a necessary condition for separability. We will see later on that partial transposition plays also an important role in establishing the distillation and activation properties of a state.

\subsection{Entanglement Witnesses}

We call an operator, $W=W^\dagger$, acting on ${\cal H}$ an $N$--partite entanglement witness (EW) (between the $N$ parties) if the following properties are fulfilled
\bi
\item[(i)] $\bra{a,\ldots, z} W\ket{a,\ldots, z} \geq 0 \quad \forall \ket{a} \in {\cal H}_A, \ldots, \ket{z} \in {\cal H}_Z$
\item[(ii)] $W$ is not positive, i.e. $W$ has at least one negative
eigenvalue
\ei

It can be easily seen that condition (i) ensures that $\tr(W \rho)\geq 0$ for any $\rho$ separable. Thus, if for some density operator, $\rho$, and an EW, $W$, $\tr(W \rho)< 0$ then $\rho$ must be entangled. In this case we say that $W$ detects $\rho$. The important point concerning separability is that a state is entangled iff there exists an EW which detects it \cite{Ho96}. 

When talking about EW's one has to distinguish two different kinds. On the one hand, there are the so--called decomposable EW (DEW), which can
be written as 
\bea
\label{decomp} 
W=O_0+O_1^{T_A}+\ldots O_N^{T_Z},
\eea
where the operators $O_i$ are positive semidefinite \cite{note00}. It can be easily verified that those witnesses cannot detect any PPTES. On the other hand non--decomposable EW (NDEW) cannot be written as (\ref{decomp}) \cite{note2}. In \cite{Le00} we showed that a $2$--partite EW, $W$, is a NDEW iff it detects a PPTES.  This result can be easily generalized to an arbitrary number of parties and so we have 

{\bf Lemma 1} A $N$--partite EW, $W$, is a NDEW iff it detects a PPTES.

\subsection{Distillation and Activation of Entanglement} 
We consider the situation where an arbitrary number of parties share an arbitrary number of copies of the state $\rho$. Then, we call $\rho$ fully distillable if the parties can, using LOCC, produce a maximally entangled state, shared among all the parties. 
Note that a PPT state can never be distilled, which can be easily seen as follows. If $\rho$ is PPT then $\rho^{\otimes N}$ is PPT for all $N$, and so is, ${\cal E}[(\rho^{\otimes N})^{T_A}]$ for all $N$ and ${\cal E}$ separable, as mentioned in Appendix B. Thus, by LOCC one can never produce, out of a PPT state a maximally entangled state, which is NPPT. Note that, in the bipartite case with $d_A=2$ and $d_B$ arbitrary, it has been shown \cite{Du99} that all NPPT states are distillable. However, for higher dimensions, there is a strong conjecture \cite {Du99,Di00} that this is no longer true, i.e. that there exist undistillable NPPT states. In the case of more than two parties, the argumentation above implies that a state can only be distillable if all the  partial transposes are non--positive semidefinite, i.e. if the state is NPPT. 

If a state is not distillable, then it might be possible to activate its entanglement using a PPTES. That is, if the parties share, in addition to $N$ copies of a state, $\rho$, a PPTES, then they might be able to produce a maximally entangled state, using LOCC. We call this process activation. Using the argumentation above, one immediately sees that only NPPT states are activable. 

Let us now, in order to be more specific, treat the case  of two parties and the one of three independently.  The next part of this section deals with the bipartite case. There we review the conditions which must be fulfilled for a state to be distillable or activable. In the second part of this section we show how to generalize the results of two parties to three.

\subsubsection{Two Parties}

In this scenario a state, $\rho\in {\cal B}({\cal H}_A\otimes {\cal H}_B)$, is called distillable if Alice and Bob can produce by LOCC a maximally entangled state (\ref{maxent}), with $d=\mbox{min}(d_{A},d_{B})$.

It has been shown \cite{Hodist} that the problem of distillation of a state, $\rho$, can be formulated in the following way:

{\bf Lemma 2} \cite{Hodist} A state, $\rho$, is distillable iff there exists a positive integer $N$ and a state of the form
\bea
\label{Schmidt2} \ket{\Psi}=
\ket{e_1,f_1}+\ket{e_2,f_2},
\eea
such that
\bea
\label{distill} \bra{\Psi} (\rho^{\otimes N})^{T_A}
\ket{\Psi} <0,
\eea
where $\{e_1,e_2\}$ ($\{f_1,f_2\}$) are two unnormalized orthogonal vectors 
in $(\C^{d_A})^{\otimes N}$ [$(\C^{d_B})^{\otimes N}$]. 

This condition simply means that iff there exists a $\C^2\otimes \C^2$ subspace, on which 
the projection of $\rho^{\otimes N}$ is NPPT, then the state is distillable. This can be 
understand as follows: if there exists such a subspace then Alice and Bob can distill a 
maximally entangled state in $\C^2\otimes\ \C^2$. They can then use some of those distilled 
states to convert them, by LOCC, into a maximally entangled state in $\C^{d_A}\otimes \C^{d_B}$. 
On the other hand, a maximally entangled state in $\C^{d_A}\otimes \C^{d_B}$ can be converted 
into a maximally entangled state in $\C^2\otimes\
\C^2$. 

We call a state $N$--distillable if for this integer $N$ condition (\ref{distill}) is fulfilled. Otherwise we call it $N$--undistillable.

If a state is not distillable, then Alice and Bob might be still able to 
distill a maximally entangled state by LOCC, if they share, in addition to their copies of the state, a PPTES. We call this process activation.

We call a $m$--undistillable state, $\rho$, activable if there exists a positive integer $N\leq m$ and a PPTES, $\sigma$, such that $\rho^{\otimes N} \otimes \sigma$ is $1$--distillable. 

Given an $N$--undistillable state, $\rho$, we call it $N$--activable if there exists a PPTES, $\sigma$, such that $\rho^{\otimes N} \otimes \sigma$ is $1$--distillable.

\subsubsection{Three Parties}

A three--partite state, $\rho\in {\cal B}({\cal H})={\cal B}({\cal H}_{A}\otimes{\cal H}_{B} \otimes{\cal H}_{C}) $ is fully--distillable if Alice, Bob and Charly can produce, using LOCC, 
out of an arbitrary number of copies of $\rho$ a GHZ--state \cite{GHZ}. Note, that this is possible 
iff (up to permutations) Alice and Bob can produce a state of the form (\ref{maxent}) with 
$d=d_{AB}=\mbox{min}(d_A,d_B)$ and Bob and Charly can distill a state of the form (\ref{maxent}) 
with $d=d_{BC}=\mbox{min}(d_B,d_C)$. This can be easily understood using that $2$ maximally 
entangled state, one in $A$ and $B$ and the other in $B$ and $C$ can always, by LOCC be combined 
to a GHZ-state, $\sum_{i=1}^{\mbox{min}(d_{AB},d_{BC})} \ket{iii}$. The other direction is also 
true, since, given two GHZ states one can, using LOCC, transform them 
into a maximally entangled state in $A$ and $B$ and one in $B$ and $C$. 
Using this fact we only have to answer the question: 'Can Alice, Bob, and Charly distill a maximally entangled state in $A,B$ and one in $B,C$?'. Therefore, from now on we will only deal with the problem of bipartite entanglement distillation. We call a state, $\rho$, $BC$--distillable if Alice, Bob and Charly can produce, using LOCC, out of an arbitrary number of copies of $\rho$ in $BC$ a maximally entangled state, i.e a state of the form (\ref{maxent}) with $d=\mbox{min}(d_B,d_C)$. Note that the best strategy for them to distill a maximally entangled state in $B$ and $C$ is that Alice measure a projector. The reason for this is that if they apply any other measurement, then they always reduce the entanglement of the outcoming state. We define $AB$, and $AC$--distillability analogously for the other cases.

If a state is undistillable we have, analogously to the bipartite case, the 
possibility to activate its entanglement using a PPTES. That is the parties 
share, in addition to some copies of their state a PPTES. Then they distill 
out of those states a maximally entangled state. If this is possible then we call the state activable. Again, we have that it is activable iff it is 
(up to permutations) $AB$--activable and  $BC$--activable; that is the entanglement among all parties can be activated iff the entanglement between $A$ and $B$ and the one between $B$ and $C$ can be activated. 

Let us now show under which conditions a state is $XY$--distillable, where $X,Y\in\{A,B,C\}$, on the example of $BC$--distillation. 

{\bf Lemma 2'} A state, $\rho$, is $BC$--distillable iff there exists a positive integer $N$ and a state of the form 
\bea 
\ket{\Psi}=\ket{e_1,f_1}+\ket{e_2,f_2}, 
\eea 
where $\{e_1,e_2\}$ ($\{f_1,f_2\}$) are two unnormalized orthogonal vectors in $({\cal H}_B)^{\otimes N}$ [$({\cal H}_C)^{\otimes N}$], and a state $\ket{h}\in ({\cal H}_A)^{\otimes N}$ such that 
\bea 
\label{abdist} 
\bra{\Psi}\bra{h} (\rho^{\otimes N})^{T_C} \ket{h}\ket{\Psi}<0. 
\eea 

Note that condition (\ref{abdist}) can only be fulfilled if $\rho$ is $B$--NPPT as well as $C$--NPPT. 

Analogously to Sec.II we call a state, $\rho$, $N$--$BC$--distillable if for the integer $N$ condition (\ref{abdist}) is fulfilled. We call $\rho$ $N$--fully distillable if it is (up to permutations) 
$N$--$AB$--distillable and $N$--$BC$--distillable.

If a state is not distillable then we still have the possibility to activate its entanglement using a PPTES. Let us now characterize those states which are $XY$--activable on the example $X=B$, $Y=C$.

We call a state, $\rho$, which is not $m$--$BC$--distillable, $BC$--activable if there exists an integer $N\leq m$ and a PPTES $\sigma$ such that $\rho^{\otimes N}\otimes \sigma$ is $1$--$BC$--distillable.

Given a $N$--$BC$--undistillable state, $\rho$, we call it 
$N$--$BC$--activable if there exists a PPTES, $\sigma$, such that $\rho^{\otimes N} \otimes \sigma$ is $1$--$BC$--distillable. We call $\rho$ $N$--fully activable if it is (up to permutations) $N$--$AB$--activable  and $N$--$BC$--activable.

\section{Characterization of Distillable and Activatable States using Entanglement Witnesses}

In this section we show that there exists a connection between EW's and the distillation and activation properties of states. We will define an operator which allows to answer the questions, if $N$ copies of a state 
are distillable, or if not, if their entanglement can be activated using a PPTES. Using this formalism it is easy to rederive the result \cite{Eg01}
that any bipartite NPPT is either $1$--distillable, or $1$--activable. One the other hand, it allows us to 
prove that there exist three--partite NPPT state which are neither $2$--distillable, nor $2$--activable. 

This section is divided into two parts. In the first part we show this connection for the bipartite case, whereas in the second part we extend those results to three parties. 
Both parts have the same structure; first we define the operator which allows us to draw the 
connection between EW's and the distillation and activation properties of a state. Then we state
our main results of the paper. In the last part of each section we prove those results. The reader,
who is not interested in the proofs can skip Sec. III A $2$ and Sec. III B $2$.

\subsection{Two Parties}

Let us denote by  $X$ is an arbitrary positive operator acting on ${\cal H}_2={\cal H}_{A_2}\otimes {\cal H}_{B_2}$. We define 
\bea 
W_{X}=P_{A_1,B_1} \otimes X^{T_{A_2}}_{A_2,B_2}, 
\eea 
where $P_{A_1,B_1}$, acting on ${\cal H}_1=\C^2\otimes \C^2$ is the projector onto the maximally entangled state (\ref{maxent}), with $d=2$. Now, we will show that $W_\rho$ allows us to answer the
questions, if the entanglement of this state can be distilled, or, if 
not, if it can be activated using a PPTES. In particular, we show that, depending on, whether $W_{\rho^{\otimes N}}$ is no EW, a NDEW, or a DEW, the corresponding state ${\rho}$ is $N$--distillable, $N$--activable, or neither $N$--distillable nor $N$--activable, respectively. This is stated by the following Theorems and Corollaries, which will be proven below.

\subsubsection{Main Results} 

{\bf Theorem 1:} A state, $\rho$, is $N$--distillable iff      
$W_{\rho^{\otimes N}}$ is no EW [it does not fulfill (i)].

{\bf Theorem 2:} A state, $\rho$, which is $N$--undistillable, is    
$N$--activable iff $W_{\rho^{\otimes N}}$ is a NDEW.                

Those theorems state that EW's do not only allow us to determine whether a state is entangled or not, but also characterize the distillability properties of a state. We have that $W_{\rho}$ is no EW iff $\rho$ is $1$--distillable (Theorem $1$).  Now, if $\rho$ is $1$--undistillable it can be either $1$--activable or not. In this case $W_{\rho}$ is an EW (Theorem $1$), which can be either decomposable or non--decomposable. Then the entanglement of $\rho$ can be activated via a PPTES iff $W_{\rho}$ is a NDEW (Theorem $2$). It is worth to mention that using these results it is not only possible to know if the entanglement of a state is distillable, or if it can be activated. It can be seen by the proofs that they provide us with a distillation protocol. That is, given a state, which can be distilled, then the separable state, which is 'detected' by the witness, $W_{\rho}$ corresponds to a LOCC, which distills a maximally entangled state (see Appendix B and \cite{Ci00}). On the other hand, the PPTES, which activates the entanglement is easily determined by the state, which is detected by the NDEW, $W_{\rho}$. Those Theorems also imply the following 
 
{\bf Corollary 1} A state, $\rho$, is is neither $N$--distillable nor $N$--activable iff $W_{\rho^{\otimes N}}$ is decomposable.

Note that the results above are not only a way of rewriting the problems, but they really allow for a new insight in the problem of distillation and activation. For instance in Sec. IV, we review in a simple manner the
fact that every bipartite NPPT state is either $1$--distillable or $1$--activable \cite{Eg01}.

\subsubsection{Proofs}
The reader who is not interested in the proofs can continue reading in the next section. For simplicity we prove the statements for $N=1$, since the argumentation holds for arbitrary $N$. The technical details and the definitions, which are needed to follow the proofs can be found in the appendix. Let us start out by determine the properties of the operator $W_X$:
\bi
\item[(a)] $W_X$ is not positive semidefinite
iff $X^{T_{A_2}}$ is not positive semidefinite,
i.e iff $X$ is NPPT.
\item[(b)] If $W_X$ fulfills condition (i) then both,
$W_X^{T_A}$ and $W_X^{T_B}$ are optimal EW's.
\item[(c)] $W_X$ is decomposable iff $W_X=R \geq 0$.
\ei

{\bf Proof:} Property (a): is clear since $P_{A_1,B_1}$ is positive. Property (b): If $W_X$ fulfills condition (i) then so do $W_X^{T_A}$ and $W_X^{T_B}$.
On the other hand $P$ is NPPT, implying that both $W_X^{T_A}$ and $W_X^{T_B}$ are not positive semidefinite and therefore they are both EW's. It remains to show that they are optimal. Using that $S_{P^{T_A}}=\{\ket{e}\ket{e^\perp} \forall \ket{e}\in \C^2\} $ (Appendix A, Proof of Lemma $4$) we find that $W_X^{T_A}$ vanishes on the set $\{\ket{e}_{A_1} \ket{e^\perp}_{B_1} \ket{\psi}, \forall \ket{e}
\in \C^2, \ket{\psi}\in {\cal H}_2\}$ (note that these are not only product states). This contains the set $S_{W_X^{T_A}}=\{\ket{e, g}_A\ket{e^\perp,f}_{B}\ \quad \forall \ket{e}\in \C^2, \forall \ket{g}\in {\cal H}_{A_2} \forall \ket{f} \in {\cal H}_{B_2}\}$. Using the fact that $\{\ket{e}\ket{e^\perp} \forall \ket{e} \in \C^2\}$ spans $\C^2\otimes \C^2$, we have that $S_{W_X^{T_A}}$ spans the whole Hilbert space, ${\cal H}=\C^2\otimes \C^2 \otimes {\cal H}_2$. Thus, $W_X^{T_A}$ is optimal. Now, using the fact that $W$ is an optimal EW iff $W^T$ is an optimal EW, we also have that $W_X^{T_B}$ is optimal. Property (c): We only have to prove the only if part. Let us assume that $W_X$ is decomposable. Then we can write it as $W_X=R+Q^{T_A}$ \cite{note00} and therefore $W_X^{T_A}=R^{T_A}+Q$. Using property (b) and the discussion concerning optimality in Appendix A we find that $Q=0$, which proves the statement $\Box$.

Note, that Property (a) tells us that, since we only have to consider NPPT states, $W_{\rho}$ is no EW iff it does not fulfill (i). With that we are now in the position to prove the results of the previous section.

{\bf Proof of Theorem 1}: Using Corollary $2$ (Appendix C) we have that $\rho$ is $1$--distillable iff there exists a separable CPM (Appendix B), ${\cal E}: {\cal B}({\cal H}_2)\rightarrow {\cal B}({\C^2\otimes \C^2})$ such that
$0>\tr_1(P_{1}^{T_B} {\cal E}(\rho))$. Using Eq. (\ref{Erho}) we can write this inequality as $0>\tr_1[P_{1}^{T_B} \tr_2(\rho^{T}_2 E_{1,2} )]= \tr_{1,2}[P_{1}^{T_B} \otimes \rho^{T}_2 E_{1,2}]$. Taking now the partial transpose with respect to $B_1,B_2$ within the trace we have $0>\tr_{1,2}[P_{1} \otimes \rho^{T_A}_2 E^{T_B}_{1,2}]= \tr(W_{\rho} E^{T_B}_{1,2})$. Thus, we have that $\rho$ is $1$--distillable iff $W_{\rho}$ 'detects' the state $E^{T_B}$. Now, since ${\cal E}$ is separable iff $E$ is separable [Appendix B (p1)], we have that the inequality above is true iff $W_{\rho}$ is no EW (since it detects $E^{T_B}$ which is separable) $\Box$. 

Note that this proofs implies the following fact. Given a distillable state, $\rho$, we determine the separable state, $E^{T_B}$, which is 'detected' by $W_{\rho}$. Then the state $E$ corresponds to the CPM, ${\cal E}$, (see Appendix B) which fulfills the property that ${\cal E}(\rho)$ is a two qubit entangled state. Thus, we found the LOCC, which distills the state, $\rho$.

{\bf Proof of Theorem 2} The proof is basically the same as the one of Theorem $1$, but now with ${\cal E}$ a PPT-preserving CPM, which implies that $E$ is PPT. Then $W$, which must be an EW (Theorem $1$), detects a PPTES and is therefore (Sec II B) a NDEW $\Box$. 

Using the same arguments as before, if we determine the PPTES, $E^{T_B}$, which is detected by the NDEW, $W_{\rho}$, then we know which PPTES activates the entanglement of $\rho$, namely $E$.

{\bf Proof of Corollary 1} $W_{\rho}$ must be either no EW, a NDEW or a DEW.
It is no EW iff $\rho$ is $1$--distillable. It is a NDEW iff $\rho$ is 
$1$--activable. Therefore $\rho$ is neither $1$--distillable nor $1$--activable iff $W_{\rho}$ is a DEW. 

\subsection{Three Parties}

Let us now generalize the results obtain in the previous section for the bipartite case to the case where we consider more parties. Here we will show how to do it for three, but one can generalize the methods introduced in the previous section to any number of particles.

Let us now denote by $X$ an arbitrary positive semidefinite operator acting on ${\cal H}_2={\cal H}_{A_2}\otimes{\cal H}_{B_2}\otimes {\cal H}_{C_2}$. We define 
\bea
\label{defW}
W_X^a =P_{B_1,C_1} \otimes X^{T_{C_2}}_{A_2,B_2,C_2} \\ 
W_X^b =P_{A_1,C_1} \otimes X^{T_{A_2}}_{A_2,B_2,C_2} \\ \nonumber
W_X^c =P_{A_1,B_1} \otimes X^{T_{B_2}}_{A_2,B_2,C_2}  \nonumber
\eea
where $P_{Y,Z}\in {\cal B}(\C^2\otimes \C^2)$ is defined in (\ref{projP}) for $\{Y,Z\} \in \{A_1,B_1,C_1\}$. 

\subsubsection{Main Results}

We will now, analogously to Sec. III C, characterize the distillation and activation properties of a state, by the operators $W_{\rho}^a, W_{\rho}^b,$ and $W_{\rho}^c$. For the seek of clarity we state our results only for the $BC$--distillation and $BC$--activation. That is we assume that $\rho$ is $B$--NPPT and $C$--NPPT (recall that otherwise it is neither possible to distill, nor to activate the entanglement shared between Bob and Charly).
All the results presented here can be formulated for the other partitions, $AB$, $AC$ too, using then the operators $W_{X}^c$ and $W_{\rho}^b$, respectively. 

{\bf Theorem 1'} A state, $\rho$, is $N$--$BC$--distillable iff $W_{\rho^{\otimes N}}^a$ is no EW [does not fulfill (i)].

{\bf Theorem 2'} A state, $\rho$, which is $N$--$BC$--undistillable, is $N$--$BC$--activable iff $W_{\rho^{\otimes N}}^a$ is a NDEW.

Those Theorems state that a state is, for instance, $1$--fully distillable iff at least two of the  operators $W_{\rho}^a, W_{\rho}^b$ or $W_{\rho}^c$ are no EW's. 
If it is not fully distillable then at least two of the operators 
$W_{\rho}^a, W_{\rho}^b$ or $W_{\rho}^c$ must be EW's (Theorem 1'). 
Then its entanglement can be fully activated via a PPTES iff at least two of the operators $W_{\rho}^a, W_{\rho}^b$ or $W_{\rho}^c$ (the one which are EW's) are NDEW's. Note that, the states which are 'detected' by $W_{\rho}^a$ are the one which allow us to derive a distillation protocol. That is, given a $1$--$BC$--distillable state, $\rho$, the separable state which is 'detected' by $W_{\rho}^a$ corresponds to a LOCC (see Appendix B), that distills $\rho$. If the state is $1$--$BC$--undistillable, but it is $1$--$BC$--activable, then the PPTES, which activates its entanglement is easily determined by the one which is detected by the NDEW $W_{\rho}^a$.  
On the other hand, we have that a state, $\rho$, is neither $1$--fully distillable nor $1$--fully activable iff at least two of the operators $W_{\rho}^a,W_{\rho}^b$ and $W_{\rho}^c$ are decomposable. 
This can be, concerning the bipartite entanglement, stated as:

{\bf Corollary 1'} A state, $\rho$, is neither $N$--$BC$--distillable nor
$N$--$BC$--activable iff there exist positive semidefinite operators $R, Q$ such that $\rho^{\otimes N}=R^{T_C}+Q^{T_B}$.

Note that, as mentioned before a three--partite state is fully distillable (activable) iff it is (up to permutations) $AB$--distillable (activable)
and $BC$--distillable (activable). Thus, Corollary $1'$ provides a necessary and sufficient condition for a state to be 
neither $N$--fully distillable nor $N$--fully activable. 

In the next section we show that using these theorems we are able to prove in a simple way that there exist three--partite NPPT states which are neither 
$1$--distillable nor $1$--activable. This is in contrast to the bipartite case, where
all NPPT states are either $1$--distillable or $1$--activable. The methods even allow us to prove that some of those states are not even  
$2$--distillable nor $2$--activable. We show that by proving that all the 
operators $W_{\rho}^a,W_{\rho}^b$ and $W_{\rho}^c$ are decomposable.

\subsubsection{Proofs}

We prove the statements above for the case $N=1$ since all the arguments remain the same for arbitrary $N$. Again, if a reader is not that interested in the proofs he can skip this part of the paper and continues reading in the next section.

We start by showing the properties of the operators $W_X^a$:
\bi
\item[(A)] $W_X^a$ is not positive semidefinite iff $X^{T_C}$ is not
positive semidefinite, i.e. X is $C$--NPPT.
\item[(B)] If $W_X^a$ fulfills condition (i) then $(W_X^a)^{T_B}$  and $(W_X^a)^{T_C}$ are optimal EW's.
\item[(C)] $W_X^a$ is decomposable iff there exists $R, Q\geq 0$ such that $X=R^{T_C}+Q^{T_B}$.
\ei

{\bf Proof} The proofs of property (A) and property (B) are similar to the ones of property (a) property (b) in Sec III A 2 and will be omitted here.
Property (C): We denote by $\tilde{Y}$ the total transpose of an operator $Y$. (if): If $X=R^{T_C}+Q^{T_B}$ then $W_X^a=P_{B_1,C_1} \otimes X^{T_{C}}=P_{B_1,C_1} \otimes(R+\tilde{Q}^{T_A})=O+S^{T_A}$, with $O=P_{B_1,C_1}\otimes R \geq 0$ and $S=P_{B_1,C_1}\otimes  \tilde{Q} \geq 0$ and so it is decomposable. (only if) If $W_X^a$ is decomposable then there exist operators (\ref{decomp}) $Q_0,Q_1,Q_2,Q_3\geq 0$ such that $W_X^a=Q_0+Q_1^{T_A}+Q_2^{T_B}+Q_3^{T_C}$ and so $(W_X^a)^{T_B}=Q_0^{T_B}+\tilde{Q}_1^{T_C}+Q_2+\tilde{Q}_3^{T_A}$ and $(W_X^a)^{T_C}=Q_0^{T_C}+\tilde{Q}_1^{T_B}+\tilde{Q}_2^{T_A}+Q_3$. Since $(W_X^a)^{T_B}$ and $(W_X^a)^{T_C}$ are both optimal EW's, because of property (B), we have that $Q_2=Q_3=0$ and so $W_X^a=Q_0+Q_1^{T_A}$. Let us now use that $P_{BC}^{T_C}$ 
and $P_{BC}^{T_B}$ are optimal decomposable EW's (Appendix A Lemma 4) and that the sets $S_{P_{BC}^{T_C}}=S_{P_{BC}^{T_B}}=\{\ket{e}\ket{e^\perp} \forall 
\ket{e} \in \C^2\}$ span $\C^2\otimes \C^2$.  This implies that $(W_X^a)^{T_C}=Q_0^{T_C}+\tilde{Q}_1^{T_B}$ vanishes on $\{\ket{e}_{B_1}\ket{e^\perp}_{C_1}\ket{\phi}_{A_2}\ket{\psi}_{B_2}\ket{\chi}_{C_2}, \forall e, \phi, \psi, \chi\}$. Thus, $Q_0^{T_C}$ and $\tilde{Q}_1^{T_B}$ must vanish on those states. This implies that $Q_0^{T_C}=P^{T_C}_{B_1,C_1}\otimes R^{T_C}$,$\tilde{Q}_1^{T_B}=P^{T_C}_{B_1,C_1}\otimes Q^{T_B}$ and so $Q_0=P_{B_1,C_1}\otimes R$,and $Q_1=P_{B_1,C_1}\otimes \tilde{Q}$ where $R,Q\geq 0$. Thus, we have that $W_X^a=P_{B_1,C_1} \otimes X^{T_{C}}=P_{B_1,C_1} \otimes(R+(\tilde{Q})^{T_A})$ and so $X=R^{T_C}+Q^{T_B}$ $\Box$.

Property (A) implies that, since we only have to consider $C$--NPPT states, 
the operator $W_{\rho}^a$ is not positive semidefinite. Thus, $W_{\rho}^a$
is no EW iff it does not fulfill (i) \cite{noteW}.

The proofs of the theorems are basically the same as the one in the previous section, 
but now with ${\cal E}:{\cal B}({\cal H}_2^{\otimes        
N})\rightarrow {\cal B}(\C^2\otimes \C^2)$, where ${\cal H}_2={\cal H}_{A_2}\otimes {\cal H}_{B_2}\otimes{\cal H}_{C_2}$
and the corresponding operator $E_{1,2}\equiv E_{A_2,B_1,B_2,C_1,C_2}\in {\cal B}({\cal H}_2^{\otimes N} \otimes \C^2\otimes \C^2)$.

{\bf Proof of Theorem 1'} Using the same arguments as in the proof of Theorem $1$ we find $\tr[P^{T_B}_{B_1,C_1} {\cal E}_a(\rho)]=\tr[W_{\rho}^a (E_{1,2}^T)^{T_C}]$, where $E_{1,2}$ is the operator corresponding to the CPM ${\cal E}_a$. Recall that ${\cal E}_a$ is separable iff $(E_{1,2}^T)^{T_C}$
is separable. Now, the right hand side of the last equation is negative, for ${\cal E}_a$ separable, iff $\rho$ is $BC$--distillable. And on the other hand, (looking at the left had side of this equation) this is true iff $W_{\rho}^a $ is no EW, since it 'detects' the separable state $(E_{1,2}^T)^{T_C}$ $\Box$.

{\bf Proof of Theorem 2'} Same Proof as for Theorem $1'$, but now
with $E$ a PPTE and ${\cal E}$ a PPT--preserving
CPM. Using that the operator $W_{\rho}^a$ must be an EW (Theorem 1'), 
we have that the PPTES, $(E^T)^{T_C}$ is detected by the EW, $W_{\rho}^a$, implying that 
it is a NDEW $\Box$.

{\bf Proof of Corollary 1'} Same Proof as for Corollary 1, but now $W_{\rho}^a$ is a DEW iff there exist operators $R, Q\geq 0$ such that $\rho=R^{T_C}+Q^{T_B}$ [property (C)] $\Box$. 

\section{Applications}

In this section we use the formalism introduced in the previous section to show the following facts. On the one hand, the entanglement of one copy of any bipartite NPPT can either be distilled or activated \cite{Eg01}. 
On the other hand, this formalism allows us to show that this is not the case for multipartite states. That is that there exist states describing a system composed of more than two subsystems which are neither $1$--distillable nor $1$--activable. Indeed, in the following subsections we will show that, using the connection between EW's and the distillation and activation properties of a state, there are NPPT states which are not even $2$--distillable nor $2$--activable.

\subsection{Two parties}

{\bf Observation} \cite{Eg01} Any bipartite NPPT state, $\rho$ is either $1$--distillable or $1$--activable.

This can be easily seen using Corollary 1, which states that $\rho$ is 
neither $1$--distillable nor $1$--activable iff $W_{\rho}$ is a DEW. Using then property (c) we have that $W_{\rho}$ is a DEW iff it is a positive semidefinite operator, which is true iff $\rho$ is PPT [property (a)].

\subsection{Example of a $1$--undistillable and $1$--unactivable three--partite state}

In this section we present a family of density operators, $\{\rho_\alpha\}$, which 
describe the state of a system composed of three qubits. 
We show, using the formalism of the previous section, that for $\alpha \leq 1$ one copy of these states can neither be distilled nor activated. Then we prove that for $\alpha \leq \alpha_0$, with $\alpha_0\approx 0.8507$
even two copies of the states can neither be distilled, nor activated.
Recall that in the bipartite case there exists no such state. 
The states of interest are 
\bea                                                
\rho_{\alpha}=\one_8+\alpha \proj{\Psi_W},                          
\eea                                                
where $\ket{\Psi_W}=\ket{001}+\ket{010}+\ket{100}$. 
Note first, that $\rho_{\alpha}$ is NPPT iff 
$\alpha >\frac{1}{\sqrt{2}}$, which implies that only in this region the state might be distillable or activable.
Note further that $\rho_{\alpha}$ is symmetric under all the permutations 
of the three parties, $A,B,C$. This symmetry implies that this state 
is $N$--$AB$--distillable iff it is $N$--$AC$--distillable iff 
it is $N$--$BC$--distillable. Let us therefore, without loos of 
generality, consider the situation where Alice performs a measurement and
Bob and Charly distill out of the remaining density operator a maximally entangled state. That is we are interested in the $N$--$BC$--distillation of the state and therefore 
in the properties of the operator $W_{\rho^{\otimes N}}^a$ Eq.(\ref{defW}). Note that, according to our definitions 
the state is (because of the symmetry) $N$--$BC$--distillable iff it is $N$--distillable. 
Then the theorems and the corollary of the previous section simplify to 

{\bf Remark 1} $\rho_{\alpha}$ is $N$--distillable iff $W_{\rho_{\alpha}^{\otimes N}}^a$
is no EW.

{\bf Remark 2} If $\rho_{\alpha}$ is $N$--undistillable it is 
$N$--activable iff  $W_{\rho_{\alpha}^{\otimes N}}^a$ is a NDEW.

{\bf Remark 3} $\rho_{\alpha}$ is neither $N$--distillable nor $N$--activable iff  
there exist positive semidefinite operators $R_{\alpha}, Q_{\alpha}$ such that 
$(\rho_{\alpha}^{\otimes N})^{T_C}=R_{\alpha}+Q_{\alpha}^{T_A}$. 

\subsubsection{One Copy}

We show that one copy of the state, $\rho_{\alpha}$ cannot be distilled for $\alpha \leq 1$, i.e (Remark $1$) $W_{\rho_{\alpha}}^a$ is an EW for $\frac{1}{\sqrt{2}}< \alpha \leq 1$. Note that the remaining state, after Alice performs a measurement is a state of two qubits, which is distillable iff it is NPPT (Sec I). And so we only have to find the measurement in $A$, $\proj{\psi}$, which maximizes the region of $\alpha$, for which the state $\bra{\psi} \rho_{\alpha} \ket{\psi}$ is NPPT. It can be easily shown that the best measurement Alice can do is to measure $\proj{0}$. Then the remaining state, $\bra 0\rho_{\alpha}\ket 0$, is NPPT iff $\alpha>1$, which implies that the state, $\rho_{\alpha}$ can be distilled, $\forall \alpha >1$. On the other hand, using Remark $1$ we have that $W_{\rho_{\alpha}}^a$ is an EW for $\frac{1}{\sqrt{2}}< \alpha \leq 1$. 
Now we show that $W_{\rho_{\alpha}}^a$ is for $\frac{1}{\sqrt{2}}<\alpha \leq 1$ a DEW and therefore $\rho$ cannot be activated for $\alpha \leq 1$ (Remark $2$). Using Remark $3$ we have to find $R_{\alpha},Q_{\alpha} \geq$ such that $\rho_{\alpha}^{T_C}=R_{\alpha}+Q_{\alpha}^{T_A}$. It can be easily verified that the operators $R_{\alpha}=\alpha \proj{\Psi^+}_{A,B}\otimes \proj{0}_C$, where $\ket{\Psi^+}=\ket{01}+\ket{10}$, and $Q_{\alpha}=\rho_{\alpha}^{T_B}-(R_{\alpha})^{T_A}$ are both positive semidefinite and lead to the desired decomposition. Thus, we have shown that the NPPT state, $\rho_{\alpha}$, is neither $1$--distillable nor $1$--activable $\forall \alpha \in ]\frac{1}{\sqrt{2}},1]$. Note that the given decomposition of $\rho_{\alpha}$ proves this statement already.
    
\subsubsection{Two copies}

Using the same method as above one can also show that two copies of $\rho_{\alpha}$ 
can neither be distilled, nor activated if $\alpha \in ]\frac{1}{\sqrt{2}},\alpha_0]$, with $\alpha_0\approx 0.8507$. In this case
$R_{\alpha}=y(\rho_{1/\sqrt{2}}^{T_C} \otimes R_1+R_1\otimes \rho_{1/\sqrt{2}}^{T_C})$, with         
$R_1=\proj{\Psi^+}_{A,B}\otimes \proj{0}_C$, where $\ket{\Psi^+}=\ket{01}+\ket{10}$, and  
$Q_{\alpha}=(\rho_{\alpha}^{\otimes 2})^{T_B}-R_{\alpha}^{T_A}$ are both positive semidefinite for $y\approx 0.4953$ and
fulfill $(\rho_{\alpha}^{\otimes 2})^{T_C}=R_{\alpha}+Q_{\alpha}^{T_A}$.

\section{Conclusions}

We have shown that on can connect the problem of entanglement witnesses to the 
one of distillation and activation of entanglement. We defined, depending on the state, $\rho$, and on the number of copies, $N$, of it an operator which has the following properties: It is an entanglement witness iff  $\rho$ is $N$--undistillable, i.e. those $N$ copies cannot be distilled via LOCC;
It is a decomposable entanglement witness iff the state is $N$--unactivatable; i.e. those $N$ copies cannot be distilled via LOCC, even if we allow for a PPTES in addition. Using those methods we have shown that there exist three--partite NPPT states, which are neither $2$--distillable, nor $2$--activable. We showed it by proving that the corresponding operator is a decomposable entanglement witness.

\section{Acknowledgments}                                        
                                                                 
This work has been supported in part by the Deutsche              
Forschungsgemeinschaft (SFB 407 and Schwerpunkt                  
"Quanteninformationsverarbeitung"), the DAAD, the Austrian       
Science Foundation (SFB ``control and measurement of coherent    
quantum systems''), the ESF PESC Programm on Quantum             
Information, TMR network ERB--FMRX--CT96--0087, the IST          
Programme EQUIP, and the Institute for Quantum Information GmbH. 

\begin{appendix}

\section{Entanglement Witnesses}

In \cite{Le00} we showed how to optimize $2$--partite EW's; that
is how to construct a new EW out of a given one,
which detects the same entangled states and in
addition some others. There we also showed that
an EW, $W$, is optimal iff $\forall
R\geq 0, \epsilon >0$ $W'=W-\epsilon R$ is not an EW, 
in the sense that it does not fulfill (i). This method of optimization can be easily 
generalized to the case of more parties. In order to recall a sufficient condition for an EW to be optimal,
we define, for an EW, $W$, the set
$S_W=\{\ket{a,\ldots z} \in {\cal H} \mbox{ such that }
\bra{a,\ldots z} W\ket{a,\ldots z}=0\}$. Then we have  

{\bf Lemma 3} If $S_W$ spans ${\cal H}$ then $W$ is
optimal.

Let us now give an example of an optimal decomposable $2$--partite EW in ${\cal B}(\C^2\otimes \C^2)$. 
This EW will allow us to draw the connection between EW's and the criterion that a state in $\C^2\otimes \C^2$
is entangled iff it is NPPT \cite{Pe96,Ho96}. 

We denote by $P_{AB}\in{\cal B}(\C^2\otimes\ \C^2)$ the projector onto the maximally entangled state, that is
\bea
\label{projP}
P_{AB}=\proj{\Phi_2}, 
\eea
with $\ket{\Phi_2}$ given in (\ref{maxent}). Then we have

{\bf Lemma 4} $P_{AB}^{T_A}$ is an optimal DEW.

{\bf Proof} Using Lemma $3$ it is sufficient to show that $S_{P_{AB}^{T_A}}$ spans $\C^2\otimes \C^2$. It can be easily verified that $S_{P_{AB}^{T_A}}=\{\ket{e, e^\perp} \forall \ket{e}\in \C^2 \}$. On the other hand on can easily check that there exists no state orthogonal to this set, which implies that $S_{P_{AB}^{T_A}}$ spans $\C^2\otimes \C^2$ $\Box$.

Note that an EW, $W$, is optimal iff $W^T$ is optimal, implying that $P_{AB}^{T_B}$ is optimal too. This EW detects, up to local operations, all NPPT states, which follows from

{\bf Lemma 5} \cite{Hodist22} A state $\rho_{AB}\in {\cal B}(\C^2\otimes \C^2)$ is NPPT iff there exist some local operators $A,B$ such that $\tr(P_{AB}^{T_B} A\otimes B \rho A^\dagger\otimes B^\dagger) <0$.

\section{Completely Positive Maps}

Any physical action can be mathematically described by a completely positive map (CPM), which is a linear, hermitian map, ${\cal E}: {\cal B}({\cal
H}) \rightarrow {\cal B}({\cal H})$, that is positive, i.e $\forall \rho \geq 0$ ${\cal E}(\rho) \geq0$, and fulfills that the extended
map, ${\cal E} \otimes \one_n$ is positive for any $n$. Note that any CPM can be written as, ${\cal E}(\rho)=\sum_k O_k \rho O_k^\dagger$,
where $O_k \in {\cal B}({\cal H})$. Note further that any separable CPM, that is any map that can be, up to a proportionality constant, implemented
locally, can be written as: 
\bea
{\cal E}(\rho)=\sum_{k=1}^l (O^A_k \otimes O^B_k
\ldots O^Z_k) \rho (O_k^A \otimes O_k^B \ldots
O_k^Z)^\dagger,
\eea
where $l$ is some finite number. Thus, we can reformulate condition (\ref{sep}) as follows; A state, $\rho$, is separable iff there exists
a separable map ${\cal E}$ such that $\rho \propto {\cal E}(\proj{00\ldots 0})$. 

A CPM, ${\cal E}$, is called $Y$--PPT--preserving if for all $\rho$ $Y$--PPT, ${\cal E}(\rho)$ is $Y$--PPT. We call a CPM PPT--preserving, if it is $Y$--PPT--preserving for all systems $Y$.

Let us also recall the isomorphism between CPM's and positive semidefinite 
operators \cite{Ci00}. We consider a CPM, ${\cal E}$, acting on the $N$ systems $A_1$,$B_1$ $\ldots$ $Z_1$ and define the operator
\bea
\label{eptoE}
E_{A_1,A_2,\ldots, Z_1,Z_2}=\frac{1}{d^{2N}}{\cal
E}(P_{A_1,A_2}\otimes \ldots P_{Z_1,Z_2}),
\eea
where $P_{X_1,X_2}$ is the projector onto the maximally entangled state (\ref{maxent}). $E_{A_1,A_2,\ldots, Z_1,Z_2}$ is acting on              
${\cal H}_A \otimes \ldots \otimes{\cal H}_Z$, with ${\cal H}_X={\cal H}_{X_1}\otimes {\cal H}_{X_2}$ and dim(${\cal H}_{X_i})=$dim(${\cal H}_{Y_i})=d$, for $i=1,2$ and $X,Y\in \{A,\ldots, Z\}$.  In Eq. (\ref{eptoE}) the map ${\cal E}$ is understood to act as the identity on the operators in ${\cal B}({\cal H}_2)$. The interpretation of Eq. (\ref{eptoE}) is the following: Each of the $N$ parties prepares his system in a maximally entangled state (locally) with an auxiliary system, e.g. $A_1$ and $A_2$ are in a maximally entangled state. Then the operation ${\cal E}$ acts on the systems $A_1,B_1,\ldots Z_1$. The state which the parties share then is proportional to $E$. On the other hand one can show that,
\bea
\label{iso1} 
&&{\cal E}(\rho_{A_1,\ldots Z_1})=\\
\nonumber &&\tr_{A_2,A_3 \ldots
Z_2,Z_3}(E_{A_1,A_2, \ldots Z_1,Z_2}
\rho_{A_3\ldots Z_3} P_{A_2,A_3}\otimes \ldots
P_{Z_2,Z_3}),
\eea
which can be also written as
\bea
\label{Erho} {\cal E}(\rho_{A1,\ldots
Z_1})=\tr_{A_2,\ldots, Z_2}(E_{A_1,A_2,\ldots
Z_2,Z_2} \rho_{A_2,\ldots Z_2}^{T})
\eea

Eq. (\ref{iso1}) has a simple physical interpretation too. If we have 
the state $E$ at our disposal we can locally implement the operation, ${\cal E}$, on a state of the systems $A_3,\ldots Z_3$ (with certain probability). For that the parties perform a joint measurement locally such that the systems $(A_2,A_3),\ldots (Z_2,Z_3)$ are projected onto a maximally entangled state (\ref{maxent}).

The importance of this isomorphism is that \cite{Ci00} 
\bi
\item[(p1)] ${\cal E}$ is a separable CPM iff $E$
is separable with respect to the systems $(A_1,A_2)
,\ldots (Z_1,Z_2)$.
\item[(p2)] ${\cal E}$ can create entanglement iff
$E$ is entangled with respect to the systems
$(A_1,A_2),\ldots (Z_1,Z_2)$.
\item[(p3)] ${\cal E}$ represents a $Y$--PPT--preserving CPM iff $E$ is $Y$--PPT. 

\ei

We can generalize this isomorphism for CPM's of the form ${\cal E}:{\cal B}({\cal H})\rightarrow {\cal B}(\bar{\cal H})$. Then the corresponding
operator $E$ is an element of ${\cal B}({\cal H}\otimes \bar{\cal H})$.

Note that (p1--p3) implies the following facts: First, we can implement a 
separable ($Y$--PPT-preserving) CPM, ${\cal E}:{\cal B}({\cal H})\rightarrow {\cal B}(\bar{\cal H})$, on a state, $\rho$, if we allow the parties to apply LOCC on the state $\rho\otimes E$, where $E\in {\cal B}({\cal H}\otimes \bar{\cal H})$ is separable ($Y$--PPT). Second, the scenario, where the parties are allowed to apply LOCC on the state $\rho\otimes E$, where $E$ is separable ($Y$--PPT), is equivalent to the one where the parties are allowed to implement a separable ($Y$--PPT-preserving) CPM.

\section{Distillation and Activation of Entanglement in terms of CPM's}

Here we use the notion of CPM's to characterize distillable and activable states. 

\subsection{Two Parties}
We consider a bipartite state, $\rho$, acting on ${\cal H}_{A_1}\otimes {\cal H}_{B_1}$.
Using Lemma $2$ and Lemma $5$ we have the following

{\bf Corollary 2} A state, $\rho$, is distillable iff there exists a positive integer $N$ and a separable CPM ${\cal E}:{\cal B}({\cal H}_1^{\otimes
N})\rightarrow {\cal B}(\C^2\otimes \C^2)$ such that
\bea
\label{distE}
\tr[P^{T_B} {\cal E}(\rho^{\otimes
N})]<0,
\eea
where $P=P_{AB}$ is given in (\ref{projP}).
Note that condition (\ref{distE}) is equivalent to $\tr[P^{T_A} {\cal E}(\rho^{\otimes N})]<0$. Note further that the operator $E$ corresponding to the CPM ${\cal E}$ is separable and acts an the Hilbert space ${\cal H}_A\otimes {\cal H}_B$, where ${\cal H}_A={\cal H}_{A_1}^{\otimes N} \otimes \C^2$ and  ${\cal H}_B={\cal H}_{B_1}^{\otimes N}\otimes \C^2$.

Using the isomorphism between density operators and CPM's (Sec II B), we know that the possibility for Alice and Bob to use a PPTES and then apply LOCC equals the possibility for them to use a PPT-preserving CPM. And so we have

{\bf Lemma 6} A state, $\rho$, which is $m$--undistillable, is activable iff there exists a positive integer $N\leq m$ and a PPT-preserving CPM ${\cal E}:{\cal B}({\cal H}_1^{\otimes N})\rightarrow {\cal B}(\C^2\otimes \C^2)$ such that $\tr[P^{T_B} {\cal E}(\rho^{\otimes N})]<0$, or, equivalently $\tr[P^{T_A} {\cal E}(\rho^{\otimes N})]<0$.

Note that the operator $E$ corresponding to the CPM ${\cal E}$ is PPT and 
acts on the Hilbert space ${\cal H}_A\otimes {\cal H}_B$, where ${\cal H}_A={\cal H}_{A_1}^{\otimes N}\otimes \C^2$ and  ${\cal H}_B={\cal H}_{B_1}^{\otimes N}\otimes \C^2$.

A $N$--undistillable state, $\rho$, is $N$--activable if for this integer $N$, condition (\ref{distE}) is fulfilled, for a PPT--preserving CPM, ${\cal E}$.

\subsection{Three Parties}

Let us now consider a density operator, $\rho$, describing the state of a system composed of three subsystems. The Hilbert Space $\rho$ is acting on is ${\cal H}_{A_1}\otimes {\cal H}_{B_1}\otimes {\cal H}_{C_1}$. Analogously to the previous section we reformulate the problem of distillation and activation in terms of CPM's. In the following we will, without loos of generality concentrate on the distillation and activation of the entanglement between Bob and Charly.

{\bf Corollary 2'} A state, $\rho$, is $BC$--distillable iff there exists a positive integer $N$ and a separable CPM  ${\cal E}_{a} :{\cal B}({\cal H}_1^{\otimes N}) \rightarrow {\cal B}(\C^2\otimes \C^2)$ such 
\bea \label{abdistE} \tr[P_{B,C}^{T_C} {\cal E}_a(\rho^{\otimes N})]<0, \eea where $P_{B,C}$ is defined in (\ref{projP}) and the small letter $a$ indicates that ${\cal E}_a$ maps a density operator describing the state of the particles $A,B,C$ to a density operator describing the state of two qubits, one held by Bob and the other one by Charly.

Then $\rho$, $N$--$BC$--distillable if for the integer $N$ condition (\ref{abdistE}) is fulfilled, for ${\cal E}_a$ separable.

{\bf Lemma 6'} A state, $\rho$, which is not $m$--$BC$--distillable, is $BC$--activable iff there exists an integer $N\leq m$ and a PPT-preserving CPM ${\cal E}_a:{\cal B}({\cal H}_1^{\otimes N}) \rightarrow  {\cal B}(\C^2\otimes \C^2)$ such that condition (\ref{abdistE}) is fulfilled. 

A $N$--$BC$--undistillable state, $\rho$, is 
$N$--$BC$--activable if for the integer $N$ condition (\ref{abdistE}) is fulfilled, for a PPT--preserving CPM, ${\cal E}_a$.

Let us summarize this section: A state is $N$--(XY) distillable (activable) iff there exists a separable (PPT--preserving) CPM, which transforms $N$ copies of the state into an entangled state acting on $\C^2\otimes \C^2$. This automatically implies that all PPTES are bound entangled, in the sense that their entanglement cannot be distilled nor activated. 

\end{appendix}



\begin{references}
\bibitem{note1}
In \cite{Di00} the problem of distillability is connected to the problem of $2$--positivity of linear maps, in contrast to the results presented here.
\bibitem{note0} Note that partial transposition depends on the used basis, but the eigenvalues do not.
\bibitem{note00} Consider, for instance,
 a bipartite DEW, $W$. It can be written, according to Eq. (\ref{decomp})
as $W=Q_0+Q_1^{T_A}+Q_2^{T_B}=Q_0+(Q_1+Q_2^T)^{T_A}=Q_0+\tilde{Q_1}^{T_A}$, with $Q_0,Q_1,Q_2\geq 0$, where we used that $Q_2^{T_B}=(Q_2^{T})^{T_A}$.
\bibitem{note2} Note that an hermitian operator can either be no EW, a DEW or a NDEW.
\bibitem{noteW} Note that $W_{\rho}^a$ does not fulfill (i) iff $(W_{\rho}^a)^{T_A}=P_{B_1,C_1}\otimes (\rho_{A_2,B_2,C_2}^T)^{T_{B_2}}$ does not fulfill (i). Thus, if $W_{\rho}^a$ does not fulfill (i) then $(W_{\rho}^a)^{T_A}$ must be in particular non--positive semidefinite. But this is the case iff $\rho^{T_B}$ is non--positive semidefinite. And so we have, if $W_{\rho}^a$ does not fulfill (i) then $\rho$ is $B$--NPPT.
\end{references}
\end{document}